\documentclass[titlepage,12pt]{article}
\usepackage[dvipdfmx]{graphicx}
\usepackage{textcomp}
\usepackage{amssymb,amsmath,color,graphics,amscd,epsf,indentfirst,amsfonts}
\usepackage{mathtools}
\usepackage{enumerate}
\usepackage{epsfig}
\usepackage{slashed}
\DeclareMathOperator{\tr}{tr}
\DeclareMathOperator{\Tr}{Tr}
\usepackage{bbm}
\newcommand{\textoverline}[1]{$\overline{\mbox{#1}}$}
\usepackage{mathabx}

\def\diag{\mathop{\rm diag}\nolimits}

\def\Tr{\mathop{\rm Tr}\nolimits}
\def\tr{\mathop{\rm tr}\nolimits}

\def\matt[#1,#2,#3,#4]{\left(%
\begin{array}{cc} #1 & #2 \\ #3 & #4 \end{array} \right)}

\def\ethe[#1,#2]{\vartheta\left[#1\atop #2\right]}

\makeatletter
\@addtoreset{equation}{section}
\@addtoreset{footnote}{page}
\makeatother

\newcommand {\slsh} [1] {\not{\hbox{\kern-2pt${#1}$}}}

\newcommand{\drawsquare}[2]{\hbox{%
\rule{#2pt}{#1pt}\hskip-#2pt
\rule{#1pt}{#2pt}\hskip-#1pt
\rule[#1pt]{#1pt}{#2pt}}\rule[#1pt]{#2pt}{#2pt}\hskip-#2pt
\rule{#2pt}{#1pt}}

\newcommand{\Yfund}{\raisebox{-.5pt}{\drawsquare{6.5}{0.4}}}

\def\drawbox#1#2{\hrule height#2pt
         \hbox{\vrule width#2pt height#1pt \kern#1pt
               \vrule width#2pt}
               \hrule height#2pt}

\def\Asym#1#2{\vcenter{\vbox{\drawbox{#1}{#2}
               \kern-#2pt       
               \drawbox{#1}{#2}}}}

\def\bdot{\huge{\textbf{.}}}

\def\beq{\begin{equation}}
\def\eeq{\end{equation}}

\numberwithin{equation}{section}

\begin{document}
\begin{titlepage} 
\vskip .7cm
\begin{center}
\font\titlerm=cmr10 scaled\magstep4
    \font\titlei=cmmi10 scaled\magstep4
    \font\titleis=cmmi7 scaled\magstep4
    \centerline{\LARGE \titlerm   GMOR relation for a QCD-like theory}
    \vskip 0.3cm
 \centerline{\LARGE \titlerm   from S-duality}
\vskip 1cm
{Adi Armoni$^\star$ and Henry Harper-Gardner$^{\natural}$}\\
\vskip 0.5cm
       {\it $^\star$ $^\natural$Department of Physics, Faculty of Science and Engineering,}\\
       {\it Swansea University, SA2 8PP, UK}\\
\vskip 0.5cm
       {\it $^\star $Center for Gravitational Physics, Yukawa Institute of Theoretical Physics,}\\
       {\it Kyoto University, Kyoto 606-8502, JAPAN}\\
\medskip
\vskip 0.5cm
{$^\star$a.armoni@swansea.ac.uk, $^\natural$2031387@swansea.ac.uk}\\

\end{center}
\vskip .5cm
\centerline{\bf Abstract}

\baselineskip 20pt
%

\vskip .5cm 
\noindent

Following \cite{Sugimoto:2012rt} we study a QCD-like gauge theory using a non-supersymmetric setup in type IIB string theory. The setup includes an $O3$ plane and $N$ D3 {\it anti}-branes and it realises a $USp(2N)$ 'electric' gauge theory with four "quarks" in the two-index antisymmetric representation and six heavy scalars in the adjoint representation. Using S-duality we obtain a dual 'magnetic' theory that includes $SO(2N-1)$ gauge theory theory with six scalars in the adjoint representation and four heavy "quarks" in the two-index symmetric representation. The dual theory provides a description of confinement and dynamical symmetry breaking of the form $SU(4)\rightarrow SO(4)$. We extend the results of \cite{Sugimoto:2012rt} by adding masses to the quarks in the electric side and deriving parts of the chiral Lagrangian using the dual magnetic theory.  In particular, we derive the Gell-Mann-Oakes-Renner (GMOR) relation for Nambu-Goldstone bosons (pions) using the magnetic theory.
 
\vfill
\noindent
\end{titlepage}\vfill\eject

\setcounter{equation}{0}

\pagestyle{empty}
\small
\vspace*{-0.7cm}

\normalsize
\pagestyle{plain}
\setcounter{page}{1}

\section{Introduction} 
\label{intro}

Confinement and chiral symmetry breaking remain the most notorious problems in QCD. There are very few analytic tools at our disposal when we consider four dimensional strongly coupled gauge theories and they usually rely on supersymmetry, where we can use the power of holomorphicity to derive non-perturbative results.

S-duality is believed to be an exact duality of type IIB string theory. When applied to a system with $N$ coincident D3 branes, the celebrated Olive-Montonen S-duality for ${\cal N}=4$ $U(N)$ SYM is recovered. Adding an O3 plane on top of the coincident branes leads to S-duality between $USp(2N)$ and $SO(2N+1)$ theories, or the SO(2N)/SO(2N) duality. These dualities and their realisation in type IIB string theory are reviewed in \cite{Giveon:1998sr}.

In ref.\cite{Uranga:1999ib} it was suggested by Uranga to study a system of O3 planes and {\it anti-}D3 branes using S-duality. A detailed study of this system was carried out by Sugimoto \cite{Sugimoto:2012rt}. In brief, the electric side of the duality provides the UV description of a QCD-like theory - a gauge theory with four flavours of massless quarks. The magnetic side provides a realisation of confinement and dynamical symmetry breaking in terms of massless 'pions' (the Nambu-Goldstone (NG) bosons). Thus the massless piece of the magnetic theory resembles the chiral Lagrangian of QCD. We review the results of \cite{Sugimoto:2012rt} in section \eqref{review}.

While the duality is interesting and the results are nice, the duality is lacking a predictive power. It turns out that the dynamics of the magnetic theory is not under full control and some assumptions needed to be made about the potential between the orientifold plane and the anti-branes. 

The purpose of the present paper is to extend and support the duality of \cite{Sugimoto:2012rt} by adding a mass to the quarks. Somewhat counter-intuitively, in QCD the mass of the pions is proportional to the square root of the sum of the quark masses, by virtue of the celebrated GMOR relation \cite{Gell-Mann:1968hlm}
\beq
f^2 _\pi M^2 _\pi = - (m_u+m_d) \langle \bar \Psi \Psi \rangle \, .
\eeq
Our main result is to recover the GMOR relation for any general set of quark masses $m_1,m_2,m_3,m_4$ by using the dual magnetic theory.

The GMOR relation is not at all natural in string theory: naively the string mass is a sum of the string tension times its length and the quark massess. How could a meson made of an open string have a mass proportional to the square root of the quarks' massess? Moreover, as we shall see, in the theory under consideration the dependence on $m_1,...,m_4$ is in general highly non-trivial and is not easily obtained from a picture of a classical string with two ends. In particular, we shall see that in general the meson mass is a function of all four quark massess.

Our paper contains several new results: we construct the chiral Lagrangian for the QCD-like theory under consideration and derive the expected mass spectrum of the Nambu-Goldstone bosons. We then derive the same mass spectrum using Sugimoto's S-duality. The matching between the QCD analysis and string theory relies on the existence of a particular scalar cubic term in the theory on the anti D3 branes and a coupling to a 3-form background flux. To our delight, all pieces fall to the right place.

The paper is organised as follows: in section \eqref{review} we review the results of \cite{Sugimoto:2012rt}. Section \eqref{chiral} is devoted to the chiral Lagrangian of the electric theory. In section \eqref{gmor} we derive the GMOR relation from the magnetic dual. Finally, in section \eqref{conclusions} we summarise our results and discuss future directions of research. 

Throughout this paper we use the following conventions: capital letters, e.g. $I,J,K=1,...,6$  are $SO(6)$ indices.  The small letters $i,j,k=1,..4$ are  used for $SO(4)$ or $SU(4)$. The small letters $a,b,c=1,2,3$ (or $\tilde a, \tilde b, \tilde c$) are used for $SO(3)$. $\alpha, \beta =1,2$ are $SU(2)$ indices. Meson masses are denoted by capital M. Quark masses are denoted by small m (or m'). Generators of Lie algebras are normalised such that ${\rm Tr}\, T^a T^b = 2\delta ^{ab}$.

\section{Review of non-supersymmetric S-duality} 
\label{review}
Before we begin to examine the dual gauge theories of interest, it is requisite to first establish the origin of their proposed duality.\\
\hfill
To begin with, we consider a type IIB string theory, in a 10-dimensional spacetime, parametrised by the usual co-ordinates \(x^0,x^1,...,x^9\). We then define a (3+1) dimensional hyperplane located at \((x^4,x^5,...,x^9)=0\), and we shall say that this plane is fixed with respect to the action of the operator \(I_6\Omega(-1)^{F_L}\), where \(I_6\) generates a \(\mathbbm{Z}_2\) action that flips the sign of the spatial co-ordinates transverse to the fixed plane \((x^{4\sim9})\), \(\Omega\) is the world-sheet parity transformation operator and \(F_L\) is the left-moving spacetime fermion number. \(I_6\Omega(-1)^{F_L}\) is called the orientifold action, and the hyperplane which is invariant under its action is called an orientifold plane, in this case it is an orientifold 3-plane, which we shall abbreviate to O3.

We construct two, seemingly distinct, string theories by placing a stack of $N$ \textoverline{D3} (anti-D3) branes in the transverse space to an \(O3^+\) or \(\widetilde{O3^-}\) plane (see \cite{Sugimoto:2012rt} for a precise definition of these orientifold planes). Note that \textoverline{D3}, like D3 branes, are invariant under the S-duality, but preserve an opposite supersymmetry to O3 planes, meaning that a system of \textoverline{D3} branes above an O3 plane breaks supersymmetry completely.  \(O3^+\) and \(\widetilde{O3^-}\)  form an S-dual pair. Therefore a stack of $N$ \textoverline{D3} branes suspended above an \(O3^+\) or \(\widetilde{O3^-}\) defines a pair of non-supersymmetric, S-dual string theories. At this point we shall implement some terminology, which will be useful for orienting our discussion moving forward. We shall refer to the theory of $N$ \textoverline{D3} branes above an \(O3^+\) plane as the 'electric side' of our duality, while the theory of $N$ \textoverline{D3} branes above an \(\widetilde{O3^-}\) will be refered to as the 'magnetic side'.\\
The Lagrangian of the electric theory at tree level is given by
\begin{align} \label{1.4}
    \mathcal{L}^{tree}_{electric}\sim\Tr{F_{\mu\nu}^2} + \Tr{(\bar{Q}_i \sigma^\mu (\partial _\mu Q^i+[A_\mu,Q^i]))} + \Tr{((\partial_\mu \Phi^I + [A_\mu,\Phi^I])^2)}\\ \nonumber
    + \Tr{(Q^i\Sigma_{ij}^I[\Phi^I,Q^j])} + \Tr{([\Phi^I,\Phi^J]^2)} + h.c. \, 
\end{align}
with $I,J=1,..,6$ and $i,j=1,..,4$.  \(A\) is the gauge field, \(Q\) is the fermion field, \(\Phi^I\) are the six transverse scalar fields, and \(\Sigma^I\) are matrices which form a Dirac-like algebra. The electric theory has the gauge symmetry \(USp(2N)\) and global symmetry \(SO(6)\). While the bosons transform in the two-index symmetric representation, the fermions transform in the two-index antisymmetric representation of the gauge group.

In line with this naming convention, we call the low-energy effective theory of the 'magnetic side' string theory, the magnetic theory, and again list its tree level Lagrangian
\begin{align}
    \mathcal{L}^{tree}_{magnetic (I)}\sim\Tr{f_{\mu\nu}^2}+\Tr{(\bar{q}_i \sigma^\mu (\partial _\mu q^i+[a_\mu,q^i]))}+\Tr{((\partial _\mu \phi^I + [a_\mu,\phi^I])^2)} \\ \nonumber
    + ((\partial _\mu + a_\mu)t)^2 + V(t) + \bar{\psi}^i\sigma^\mu(\partial _\mu \psi_i + a_\mu \psi_i)  + \Tr{(q^i\Sigma_{ij}^I[\phi^I,q^j])}\\ \nonumber
    + \Tr{([\phi^I,\phi^J]^2)} + t^T\phi^I\phi^It + \bar{\psi}^{iT}\Sigma^I_{ij}\phi^I\bar{\psi}^j + t^Tq^i\psi_i + h.c.
\end{align}

Where \(a\) is the gauge field, \(q\) and \(\psi\) are the fermions, \(t\) is a tachyonic scalar field and again \(\phi^I\) are six tranverse scalar fields. The magnetic theory has gauge symmetry \(SO(2N)\) and global symmetry \(SO(6)\). Note that the gauge field and the scalars $\phi ^I$ transform in the antisymmetric representation, while the fermions $q^i$ transform in the symmetric representation of the gauge group. The tachyon $t$ and the fermion $\psi$ transform in the fundamental representation.

The electric and magnetic theory Lagrangians are in schematic form, and the couplings have been omitted. There are some issues to clarify before any productive analyses of these theories can proceed. Starting with the most obvious point of concern, we have a tachyon in our magnetic theory, making it unstable. After tachyon condensation the gauge symmetry will be broken to \(SO(2N) \rightarrow SO(2N-1)\), and through the terms \(t^T\phi^I\phi^It\) and \(t^Tq^i\psi_i\) we see that this condensation will give mass to some components of \(\phi^I\) and \(q^i\) as well as to $\psi$. The massless field Lagrangian after tachyon condensation is given here
\begin{align} \label{1.6}
    \mathcal{L}^{tree}_{magnetic (II)}\sim\Tr{f_{\mu\nu}^2}+\Tr{(\bar{q}_i \sigma^\mu (\partial _\mu q^i+[a_\mu,q^i]))}+\Tr{((\partial _\mu \phi^I + [a_\mu,\phi^I])^2)} \\ \nonumber
     + \Tr{(q^i\Sigma_{ij}^I[\phi^I,q^j])}
    + \Tr{([\phi^I,\phi^J]^2)} + h.c.
\end{align}
Going forward, this is the theory we will be referring to as the magnetic theory. It is similar to ${\cal N}=4$ $SO(2N-1)$ SYM, except that the fermions $q^i$ transform in the two-index symmetric representation of the gauge group.  \\
\hfill

 The electric theory is asymptotically free, in other words, it is best described at high energies. The fields in \eqref{1.4} are all massless at tree level, however as the supersymmetry is completely broken, the scalar fields take on cut-off scale masses from the quantum corrections and decouple from the low energy theory. It is conjectured in the literature that the \(USp(2N)\) gauge theory with four Weyl fermions in the anti-symmetric representation lies outside the conformal window and is in the confined phase  \cite{Sannino:2009aw,Armoni:2009jn}. In particular, it was argued \cite{Armoni:2009jn} that the critical number of flavours above which $USp(2N)$ theory with antisymmetric fermions is in the conformal window is $N_{f}^{\star} = 4\frac{N_c+1}{N_c-1}$. In addition, for the case that \(N>1\) the global \(SO(6)\) symmetry is believed to be dynamically broken to \(SO(4)\) by the condensation of a fermion bilinear
\beq
 \epsilon^{\alpha\beta}\langle\Tr{(Q^{i}_\alpha Q^{j}_\beta)}\rangle =c \delta^{ij} \, , \label{cond}
\eeq
with $c$ the value of the quark condensate.

The magnetic theory is more opaque. It is asymptotically non-free, which raises obstacles when taking a decoupling limit in a controlled way, however, some useful insight is extracted from a comparison of 1-loop corrections to the scalar masses between the electric and magnetic theories. The 1-loop calculation shows that the mass-squared for the \(\Phi^I\) in the electric theory is positive, while the mass-squared for the \(\phi^I\) in the magnetic theory is negative
\begin{equation} \label{1.7}
    m_{\Phi}^2=+Cg_sl_{s}^{-2} \textrm{, } m_{\phi}^2=-C'g_sl_{s}^{-2}
\end{equation}

Where \(g_s\) is the string coupling and \(C\) and \(C'\) are positive constants.\\

The scalar fields on the \textoverline{D3} brane have the interpretation of the position of the brane in the transverse space to the central O3 plane, as such, the mass-squared values in \eqref{1.7} suggest that the \textoverline{D3} in the electric theory are attracted to the O3 plane, while in the magnetic theory they are repulsed. 

 With this interpretation, \eqref{1.7} gives us a picture of a scalar theory which is unstable at the origin, but becomes attractive towards the origin at large distances. This Higgs potential-like behaviour suggests that our scalar fields will develop a non-trivial expectation value (vev) and will thereby spontaneously break the global symmetry of the system. This qualitative picture is of course best described mathematically on the magnetic side of our duality, in spite of the difficulties concerning the energy scales at which it is strongly or weakly coupled, because the scalar fields are completely decoupled from the electric theory. Rather than going through the hardship of attempting to calculate the exact potential of the magnetic theory, Sugimoto proposed a 'toy' potential, or model, for the magnetic theory scalars
\begin{equation}\label{1.8}
    V(\phi^I)=-\frac{\mu^2}{2}\tr(\phi^I\phi^I)-\frac{g}{4}\tr([\phi^I,\phi^J]^2)+\frac{\lambda}{2}\tr((\phi^I\phi^I)^2)
\end{equation}
Where the first term is the tachyonic mass term, the second is imported from the potential portion of the Lagrangian \eqref{1.6}, and the final, quartic term is included to stabilise the potential at long-distance, to reflect the behaviour we know to expect. It should be noted that the last term is expected to be generated in a non-supersymmetric theory, according to the general rules of QFT. Any operator that preserves the symmetries of the tree level Lagrangian can be generated via quantum corrections. The sign and the magnitude of $\lambda$, however, were assumed in \cite{Sugimoto:2012rt}, in order to ensure a vacuum with symmetry breaking.


Differentiating \eqref{1.8} yields the following equation of motion for the magnetic theory
\begin{equation} \label{1.9}
     -\mu^2\phi^I -g[\phi^J,[\phi^I,\phi^J]] +\lambda(\phi^I(\phi^J\phi^J)+(\phi^J\phi^J)\phi^I)=0
\end{equation}\\
Which admits several vacua, depending on the choice of (in)equality between the positive coefficients \(\lambda\) and \(g\).

The scalar field \(\phi^I\) takes values in the Lie algebra of the gauge group. For the case $N=2$ the group is \(SO(3)\), hence we define
\begin{equation}
    \phi^I=A^I_iJ^i \, ,
\end{equation}
where \(J^i\) are basis elements of the Lie algebra \(so(3)\) (the spin-1 representation of \(su(2)\)). For the choice \(\lambda>g\), it is straightforward to show that the following value for \(\phi^I\) is a solution to \eqref{1.9} and thereby a vacuum of the theory
\begin{equation} \label{2.11}
    \langle \phi^1 \rangle =aJ^1, \langle \phi^2 \rangle =aJ^2, \langle \phi^3 \rangle =aJ^3, \langle \phi^{4\sim6} \rangle=0 \, .
\end{equation}
This is a ``fuzzy sphere'' configuration of radius $a$. This solution is chosen as it is the only admissible vacuum that breaks $SO(6)\rightarrow SO(3)\times SO(3)$. This is sufficient to justify the choice as it is assumed that the $SL(2,\mathbbm{Z})$ symmetry which generates the duality is exact, this being so, the electric and magnetic theories must exhibit the same symmetry breaking pattern. As we shall see later, the constant $a$ corresponds to the value of the quark condensate.

This vacuum for \eqref{1.8} is clearly invariant under the group \(SO(3)\times SO(3)\), where one SO(3) acts on the non-zero components of \(\langle\phi^I\rangle\) and is freely undone by a gauge rotation, and the other is the SO(3) which acts on the null-components of \(\langle\phi^I\rangle\) which are trivially invariant under its action. The isomorphism \(SO(3)\times SO(3) \simeq SO(4)\) is a known result. Therefore, the duality allows us to realise the dynamical \(SO(6) \rightarrow SO(4)\) symmetry breaking expected of the strongly-coupled electric theory very simply and elegantly in terms of the condensation of a non-zero vev for scalar fields in its magnetic-dual description.

A careful study of the vacuum \eqref{2.11}, carried out in \cite{Sugimoto:2012rt}, reveals that all the fields, except nine modes which transform in the coset  $SO(6)/SO(4)$ become massive, as expected from the Goldstone theorem. In particular, the gauge bosons acquire a mass due to SSB and the fermions acquire a mass due to Yukawa couplings.  Moreover all the scalars, except the nine modes that correspond to the NG bosons, also acquire a mass.

\section{The chiral Lagrangian} 
\label{chiral}

In a $USp(2N)$ theory with four antisymmetric quarks there exists a $U(4)=U(1)\otimes SU(4)$ global symmetry. The $U(1)$ part is anomalous and hence the theory admits a massive $\eta'$ pseudo-scalar meson. According to Witten-Veneziano formula \cite{Witten:1979vv,Veneziano:1979ec} we expect it to have a mass $M^2_{\rm \eta'}\sim \frac{2N-2}{2N}\Lambda _{\rm QCD}^2$.  Unlike ordinary QCD where the $\eta'$ becomes light in the 't Hooft large-$N$ limit, in the present case the $\eta '$ is always heavy and therefore it decouples from the low-energy theory.

The global $SU(4)$ symmetry is expected to break dynamically according to the pattern 
\beq
  SU(4) \rightarrow SO(4).
\eeq

The order parameter for the breaking is the quark condensate \eqref{cond}.

The breaking of the global symmetry results in a multiplet of nine massless Nambu-Goldstone (NG) bosons. The NG bosons belong to the coset $U\equiv G/H=SU(4)/SO(4)$. The fifteen generators of the $SU(4)$ are either symmetric (and real) or antisymmetric (and imaginary) Hermitian matrices. The six antisymmetric generators form the generators of the $SO(4)$ group.  The remaning nine symmetric generators of the $SU(4)$ group transform in the two-index traceless symmetric representation of $SO(4)$. The massive $\eta'$ particle corresponds to the unit matrix.

 The chiral Lagrangian of the present theory can be written in terms of $U$, with 
\beq
U=\exp i\pi \, ,
\eeq
where $\pi$ is a matrix that transforms in the two-index traceless symmetric representation of the $SO(4)$ algebra.

The relevant terms that will be at the centre of our interest are the kinetic term and the mass term for the NG bosons (the 'pions')
\beq
S \sim \int d^4 x \,  {\rm Tr} \left ( (U^{-1} \partial _ \mu U) (U^{-1} \partial  ^\mu U) + c( m_q U + {\rm h.c.}) \right ) \,  ,
\eeq
 where $m_q$ is the quarks' mass matrix, namely the same $4\times 4$ symmetric matrix that gives mass to the quarks
\beq
 (m_q)_{ij} Q^i Q^j + {\rm h.c} .
\eeq
We will choose $m_q$ to be the diagonal matrix $m_q={\rm diag}(m_1,m_2,m_3, m_4)$. Note that we also set $f_\pi =1$.

We will mostly be interested in the kinetic term and the mass term of the pions
\beq
S = \int d^4 x \,  \frac{1}{2} {\rm Tr} \left ( \partial _\mu \pi \partial ^\mu \pi - c m_q \pi ^2\right )+... \,  . \label{pions}
\eeq
In the simplest case, where all four quarks have the same mass $m_q=m \mathbbm{1}$, we recover the celebrated GMOR relation
\beq
M^2 _\pi \sim c m\, .
\eeq

In the most general case where the four quark masses have arbitrary values we can proceed as follows. We parametrize the symmetric $4\times 4$ pion matrix using ten entries, such that $\pi _{ij}= \pi _{ji}$. Note that the diagonal is not traceless, namely we have ten Nambu-Goldstone bosons instead of nine. We thus add the constraint
\beq \sum _i \pi _{ii}=0 \, . \label{constraint} \eeq 

The mass terms in \eqref{pions} together with the constraint \eqref{constraint} take the form
\beq
{\cal L}= 
  -\frac {c}{2} \sum_{ij} (m_i+m_j) \pi _{ij}^2 - \Lambda ^2 (\sum _i \pi_{ii})^2  \, ,
\eeq

with $\Lambda \rightarrow \infty$. We may think about $\sum _i \pi_{ii}$ as an infinitely heavy $\eta'$. If, instead, we consider a hypothetical theory where $\Lambda =0$, namely we ignore the constraint \eqref{constraint}, we obtain at low eneregy ten light particles whose masses are given by $M_{ij}^2=c(m_i + m_j)$, where four of them contain a quark anti-quark pair of same flavour ($m_i=m_j$)  and the other six contain a quark anti-quark of different flavours ($m_i \ne m_j$). Imagine that we continuously vary the value of $\Lambda$ from $0$ to $\infty$. As we increase $\Lambda$ the mass of the $\eta'$ increases, the masses of six NG bosons do not change and the mass of the three remaining NG bosons become a mixture of the four quark masses. The precise eigenvalues are determined by diagonalising a $3\times 3$ matrix. We will discuss it in more detail in the next section.

At the Lie Algebra level $so(4)$ is isomorphic to $so(3)\times so(3)$. For comparison with the results of S-duality, it will be more convenient to write the chiral Lagrangian in the language of $SO(3)\times SO(3)$. The nine pions which transform in the traceless symmetric representation of $SO(4)$ transform in the bi-fundamental of $SO(3)\times SO(3)$. The ten entries of the symmetric mass matrix $m_{ij}$ can be decomposed into a singlet $m'$ and nine bifundamentals ${m'}_{\tilde a}^a$ of $SO(3)\times SO(3)$ ($a,{\tilde a}=1..3$), as listed in table \eqref{tchiral} below
\begin{table}[!ht]
\begin{center}
\begin{tabular}{|c|cc|}
\hline
 & $SO(3)$ & $SO(3)$ \\
\hline
 $m'$ & \bdot & \bdot  \\
${m'}_{\tilde a}^a$ & \Yfund & \Yfund \\
\hline
$\pi_{\tilde a}^a$ & \Yfund & \Yfund \\               
\hline
\end{tabular}
\caption{\it  Content of the chiral Lagrangian.}
\label{tchiral}
\end{center}
\end{table}
\\
The relation between the ten parameters of $(m_q)_{ij}$ and $m',{m'}_{\tilde a}^a$ is given in appendix \eqref{relation}.

The explicit form of the action \eqref{pions} is
\beq
S = \int d^4 x \,  \frac{1}{2} \left (  \partial _\mu \pi _{\tilde a}^a \partial ^\mu \pi ^{a}_{\tilde a} - c m'  \pi _{\tilde a}^a  \pi ^{a}_{\tilde a} + c \epsilon ^{\tilde a \tilde b \tilde c} \epsilon _{abc} \tilde m{'} _{\tilde a}^{a}   \pi _{\tilde b}^{b}  \pi _{\tilde c}^{c} \right ) . \label{pions2}
\eeq
We will derive eq.\eqref{pions2} in the next section.

\section{GMOR relations from S-duality} 
\label{gmor}

Let us introduce mass to the quarks of the electric theory and examine how it affects the mass of the pions in the magnetic side of the duality. To this end we will introduce a three-form flux $G_3 = F_3 - \tau H_3$ in the type IIB background following \cite{Camara:2003ku,Polchinski:2000uf}. $F_3$ and $H_3$ are RR and NSNS three-form fluxes. As we shall see in a moment the flux encodes the quark mass matrix $(m_q)_{ij}$.

The action of a D3 brane in a background that includes a three-form flux is given by \cite{Camara:2003ku} and contains the following terms.
\beq \label{4.1}
{\cal L}_{\rm soft} = ... + i\frac{g_s}{6}(\star _6 G_3 - iG_3) _{IJK} \phi ^I \phi ^J \phi ^K +  i\frac{g_s}{96}(\star _6 G_3 - iG_3) _{IJK}Q \gamma ^{[I} \gamma ^J \gamma ^{K]} Q + {\rm h.c.}
\eeq

Substituting the following components for the three-form term. \begin{equation}
   (\star _6 G_3 - iG_3) \label{4.2} _{IJK}=C_{IJK}=\frac{-1}{48}\Tr(m_q(\epsilon_{IJK}^{I'J'K'}\gamma_{[I'}\gamma_{J'}\gamma_{K']}- i\gamma_{[I}\gamma_J\gamma_{K]})) 
\end{equation}
which we derive in the appendix, we find that the three-form coupling confers a fermion mass term to the electric theory of the form \(Q^i(m_q)_{ij}Q^j\) where we have full control over the entries of the matrix \(m_q\).

The reader will recall, from section (2), that in \cite{Sugimoto:2012rt} the scalar fields of the electric theory acquire cut-off scale masses and decouple. Consequently, only the flux-induced quark mass term of \eqref{4.1} carries into the electric theory. However, in the magnetic theory, the scalar fields are where the critical behaviours of the physics are realised, and as such only the scalar coupling term of \eqref{4.1} is of interest to the magnetic theory. Our specific aim is to relate the quark masses and pion masses due to this three-form coupling, therefore we introduce the scalar coupling in \eqref{4.1} as a perturbation around the fuzzy sphere vacuum \eqref{2.11} of the original magnetic theory potential.

\begin{equation} \label{4.3}
    V'=V_0 +\frac{ig_s}{6}\Tr(C_{IJK}\phi^I\phi^J\phi^K)
\end{equation}
Where \(V_0\) is the potential \eqref{1.8}. The trace is over the gauge group of the magentic theory $SO(2N-1)$. For simplicity we will consider the case with $N=2$, namely $SO(3)$. The generalisation to arbitrary $N$ is straightforward.

We introduce perturbations to the scalar fields of the following form
\begin{equation}
    \phi^I = \langle\phi^I \rangle +\delta\phi^I
\end{equation}

where \(\delta\phi^{1\sim 3}=0,\delta\phi^{I=4 \sim 6}=\delta A^I_aJ^a\).

Substituting this into the three-form coupling term in \eqref{4.3} we obtain several terms of varying order in the vev and perturbations, but most important for our purposes is the following
\begin{align} \label{4.5}
    \Tr{C_{IJK}(\langle\phi^I\rangle+\delta\phi^I)(\langle\phi^J\rangle+\delta\phi^J)(\langle\phi^K\rangle+\delta\phi^K)}=\\
    \nonumber
    ...+3\Tr{C_{IJK}(\langle\phi^I\rangle\delta\phi^J\delta\phi^K)+...}
\end{align}

The terms we omit include those which are linear in the scalar perturbation, these are 'tadpoles' in the language of QFT, and do not contribute to the dynamics. There are also terms which are cubic in the perturbation, while these do contribute to the dynamics of the perturbed fields they are interaction terms, and are not relevant to questions about the pion masses. 
We find then that this is a mass term for the pions, as they are quadratic in the perturbed scalar fields. We know also, by construction, that the background three-form \(C\) is linear in the quark masses, what emerges then, is that the quark and pion masses are related schematically as \(M_\pi^2\sim m\). This result is known from the chiral Lagrangian of QCD and is often referred to as the Gell-Mann, Oakes, Renner (GMOR) relation. We wish to go further than a schematic comparison however, rather we aim to extract the full, general relationship between the quark and pion masses under the duality.

To proceed we examine \eqref{4.5} in more detail. Note that the trace over the scalar fields in the magnetic theory potential is a trace over \(SO(3)\), namely the gauge group of the magnetic theory. While the three-form \(C\) contains a trace over \(SO(4)\). So in \eqref{4.5} we are dealing with a term which involves a nested trace over two different groups.
\begin{align}
    3\Tr_{SO(3)} (C_{IJK}\langle\phi^I\rangle\delta\phi^J\delta\phi^K)
    =\frac{-1}{16}\Tr_{SO(3)}( \Tr_{SO(4)}(m_q(\epsilon_{IJK}^{I'J'K'}\gamma_{[I'}\gamma_{J'}\gamma_{K']}- i\gamma_{[I}\gamma_J\gamma_{K]}))\langle \phi^I \rangle \delta\phi^J \delta\phi^K)
\end{align}
Note, in our expression for the components of the background three-form \(C_{IJK}\), we are free to use our non-chiral expression which is derived in the appendix, as the first portion which contracts with the epsilon is not significant to calculations concerning the pion masses. With this in mind, we can greatly simplify our term.
\begin{align} \label{derivedpionmass}
   3\Tr_{SO(3)} (C_{IJK}\langle\phi^I\rangle\delta\phi^J\delta\phi^K)
    =\frac{-1}{8}\Tr_{SO(3)}(\Tr_{SO(4)}(m_q\gamma_{[I}\gamma_J\gamma_{K]})\langle \phi^I \rangle \delta\phi^J \delta\phi^K)
\end{align}

The nested trace of \eqref{derivedpionmass} is a bar to progress, as it mixes the \(SO(4)\) language of the electric theory with the \(SO(3)\) language of the gauge symmetry of the magnetic theory. However, we can resolve this by reconsidering the space our pions live in. Our magnetic theory pions are Nambu-Goldstone bosons, which take values as the generators of the coset \((SO(6)/(SO(3) \times SO(3)))\), associated with the dynamical symmetry breaking. We can show that in this case the Nambu-Goldstone bosons transform under the symmetry group of the vacuum, and as such the trace taken over \eqref{derivedpionmass} should be a trace over the bifundamental representation space of \(SO(3) \times SO(3)\). This is achieved by making a 'colour-flavour' identification, that is, we identify the \(SO(3)\) gauge symmetry of the magnetic theory with one of the copies of \(SO(3)\) which lives in the \(SO(4)\) symmetry of the electric theory (recall the isomporphism \(SO(4) \simeq SO(3)\times SO(3)\)). i.e.
\begin{equation*}
    SO(3)_{\textrm{col.}}\sim SO(3)_{\textrm{flav.}}
\end{equation*}
\begin{equation*}
    \implies \Tr_{SO(3)}\Tr_{SO(4)} \rightarrow \Tr_{SO(3) \times SO(3)}
\end{equation*}

Of course in order to perform an \(SO(3) \times SO(3)\) trace over the pion mass term, all the factors must be in the bi-fundamental representation of \(SO(3) \times SO(3)\). However, from the outset, \(m_q\) and the Dirac matrices \(\gamma^I\) belong to a representation space of \(SO(4)\). \(m_q\) is a real, symmetric, \(4\times 4\) matrix, which we may view as having two components, a traceful, and a traceless. Both of these parts may be treated as representation spaces of \(SO(4)\). The traceless part of \(m_q\) is a nine dimensional representation with the \(SO(4)\) action \(m \rightarrow OmO\), where \(m\in\) traceless, symm. Mat(\(4,\mathbb{R}\)). The traceful component of \(m_q\) is the one dimensional 'singlet' representation of \(SO(4)\). Furthermore, in \cite{Sugimoto:2012rt} representation of the Dirac algebra, the elements of which we have labelled \(\gamma^I\) are generators of \(SO(4)\), which can in turn be viewed as a double copy of the Lie algebra of \(SO(3)\).

To arrive at a mass term for the pions which can be fully evaluated in the language which is the natural to the magnetic theory moduli space, we must map the traceful and traceless components of \(m_q\) from their respective \(SO(4)\) the appropriate representations of \(SO(3)\times SO(3)\). To be explicit, we wish to map,
\begin{equation} \label{map}
    (SO(4)): m\mathbbm{1}_{4\times 4}+m^\mu T_\mu \rightarrow (SO(3)\times SO(3)): m' 1 \otimes 1 +m'^{\; a}_{\; \tilde{a}} J_a\otimes J^{\tilde{a}}
\end{equation}
Where \(T_\mu\) (with \(\mu=1,...,9\)), are a basis of traceless, symmetric Mat(\(4,\mathbbm{R}\)). \(J^a,J_{\tilde{a}}\) (with \(a,\tilde{a}=1,2,3\)), are generators of the Lie algebra \(so(3)\).\\
\hfill

We are only concerned with mapping \(m_q\) of the form \(m_q=\diag(m_1,m_2,m_3,m_4)\), as this corresponds to the most general quark mass term in our electric theory. Therefore we may decompose \(m_q\) to the form given on the left-hand side of \eqref{map} as follows
\begin{align} \label{decomposition}
    m_q=\frac{1}{4}(m_1+m_2+m_3+m_4)\mathbbm{1}+\frac{1}{4}(m_1+m_2-m_3-m_4)\diag(1,1,-1,-1)\\ \nonumber
    +\frac{1}{4}(m_1+m_4-m_2-m_3)\diag(1,-1,-1,1)+\frac{1}{4}(m_2+m_4-m_1-m_3)\diag(-1,1,-1,1)
\end{align}

The pion mass term \eqref{derivedpionmass} may then be expressed in a form which makes the realisation of the map \eqref{map} straightforward.
\begin{equation} \label{masstermso4}
    \Tr(M_\pi^2\delta\phi\delta\phi)=\frac{-1}{8}\Tr((m\mathbbm{1}_{4\times 4}+m^{\mu} T_{\mu}){\langle \phi^I \rangle}\gamma_I \delta\phi^J \gamma_J \delta\phi^K \gamma_K)
\end{equation}
Note that \(m=(\frac{m_1+m_2+m_3+m_4}{4})\) and \(m^{\mu}\) has non-zero components \((\frac{m_1+m_2-m_3-m_4}{4})\), \((\frac{m_1+m_4-m_2-m_3}{4})\), \((\frac{m_2+m_4-m_1-m_3}{4})\).

We now implement the isomorphism \(SO(3)\times SO(3) \simeq SO(4)\). Naturally, the tracefull part of \(SO(4)\) is mapped to the singlet of \(SO(3)\times SO(3)\). The traceless, diagonal matrices of \eqref{decomposition} (\(T_{\mu}\)), are mapped to elements of the bifundamental algebra of \(SO(3)\times SO(3)\). The Dirac matrices (\(\gamma_a\)) are mapped to basis elements (\(J_a\)) of the Lie algebra \(so(3)\). Finally, the factor of \(\langle \phi^I \rangle\gamma_I\), which the singlet of the vacuum symmetry, is mapped to the singlet of \(SO(3)\otimes SO(3)\), multiplied by the constant (\(a\)) which we associate with the fermion bilinear condensate. To summarise the map, we tabulate the transformation of each factor in \eqref{masstermso4} below

\begin{table}[!ht]
\begin{center}
\begin{tabular}{|c|c|}
\hline
 $SO(4)$ & $SO(3)\times SO(3)$ \\
\hline
 $m\mathbbm{1}$ & \(2m' 1\otimes 1\) \\
\hline
$T^{\mu}$ & \(J_a\otimes J^{\tilde{a}}\) \\
\hline
\(m^{\mu}\) & \({m'}_{\; \tilde{a}}^{\; a}\) \\
\hline
\(\langle \phi^I \rangle\gamma_I\) & \(a 1\otimes 1\) \\
\hline
\end{tabular}
\label{SO4toSO3xSO3map}
\end{center}
\end{table}

Note that the isomorphism \(SO(3)\times SO(3) \simeq SO(4)\) is actually an isomorphism at the level of the Lie algebras of both groups. As such we are at liberty to map \(T_{\mu} \rightarrow J_a\otimes J^{\tilde{a}}\) in whatever way is convenient. For reasons of neatness further along in the process, we have chosen (\(T_1\rightarrow J_1\otimes J^{\tilde{1}}, T_2\rightarrow J_2\otimes J^{\tilde{2}}, T_3\rightarrow J_3\otimes J^{\tilde{3}}\)). This gives (\({m'}_{\; \tilde{a}}^{\; a}\)) as follows

\begin{equation}
    {m'}_{\; \tilde{a}}^{\; a} =\begin{pmatrix}
    \frac{m_1+m_2-m_3-m_4}{4} & 0 & 0\\
    0 & \frac{m_1+m_4-m_2-m_3}{4} & 0\\
    0 & 0 & \frac{m_2+m_4-m_1-m_3}{4}\\
    \end{pmatrix}
\end{equation}

and in addition $m'=m=\frac{1}{4}(m_1+m_2+m_3+m_4)$, see appendix B.

We can now express the pion mass term fully in terms of the \(SO(3)\times SO(3)\) language.
\begin{align}
    \Tr(M_\pi^2\delta\phi\delta\phi)=\frac{-a}{8}\Tr((2m' 1\otimes 1 +m'^{\; a}_{\; \tilde{a}} J_a\otimes J^{\tilde{a}})(\delta A^b_{\tilde{b}}J^{\tilde{b}}\otimes J_b )(\delta A^c_{\tilde{c}}J^{\tilde{c}}\otimes J_c))
\end{align}

There is subtlety we need to address: the matrix $m_q$ acts on Dirac fermions. For this reason the group we need to consider is actually the group $SU(2)\times SU(2)$ and we therefore choose $J_i \equiv \sigma _i$, the Pauli matrices. Note that \(\Tr(J_aJ_b)=2\delta_{ab}\), \(\Tr(J_aJ_bJ_c)=2i\epsilon_{abc}\).

We can now begin to explicitly evaluate our pion mass term.
\begin{align}
    \Tr(m'(\delta A^b_{\tilde{b}}J^{\tilde{b}}\otimes J_b )(\delta A^c_{\tilde{c}}J^{\tilde{c}}\otimes J_c))=m'\delta A^b_{\tilde{b}}\delta A^c_{\tilde{c}}\Tr(J^{\tilde{b}}\otimes J_b J^{\tilde{c}}\otimes J_c)\\ \nonumber
    =4m'\delta A^b_{\tilde{b}}\delta A^c_{\tilde{c}}\delta^{\tilde{b}\tilde{c}}\delta_{bc}
\end{align}

\begin{align} \label{massterm3}
    \Tr((m'^{\; a}_{\; \tilde{a}} J_a\otimes J^{\tilde{a}})(\delta A^b_{\tilde{b}}J^{\tilde{b}}\otimes J_b )(\delta A^c_{\tilde{c}}J^{\tilde{c}}\otimes J_c))=m'^{\; a}_{\; \tilde{a}} \delta A^b_{\tilde{b}}\delta A^c_{\tilde{c}}\Tr(J_a\otimes J^{\tilde{a}})(J^{\tilde{b}}\otimes J_b)( J^{\tilde{c}}\otimes J_c)\\ \nonumber
    =-4m'^{\; a}_{\; \tilde{a}}\delta A^b_{\tilde{b}}\delta A^c_{\tilde{c}}\epsilon^{\tilde{a}\tilde{b}\tilde{c}} \epsilon_{abc}
\end{align}

Our full mass term for the magnetic theory pions is then,
\begin{align} \label{massterm4}
    M_\pi^2 \delta A^a_{\tilde{a}}\delta A_{\tilde{a}}^a=\frac{-a}{8}(8m'\delta A^a_{\tilde{a}}\delta A_{\tilde{a}}^a-4{m'}^{\; a}_{\; \tilde{a}}\delta A^b_{\tilde{b}}\delta A^c_{\tilde{c}}\epsilon^{\tilde{a}\tilde{b}\tilde{c}} \epsilon_{abc})
\end{align}

We see that this is the pion mass term derived from the chiral Lagrangian in \eqref{pions2}. The radius of the fuzzy sphere, namely the constant $a$, is identified with the value of the quark condensate $c$ in field theory.

However, \eqref{massterm4} is not yet an entirely sensible mass-squared term for our pions, as is apparent when we contract the indices on the right-hand side of the equation. The first portion is well-behaved as we get nine terms of the form \(m'(\delta A^{\tilde{a}}_a)^2\), but a problem occurs in the second portion.

\begin{align} \label{PionMassExpanded}
    m'^{\; a}_{\; \tilde{a}}\delta A^b_{\tilde{b}}\delta A^c_{\tilde{c}}\epsilon^{\tilde{a}\tilde{b}\tilde{c}} \epsilon_{abc}=m'^{\; 1}_{\; \tilde{1}}(\delta A^2_{\tilde{2}}\delta A^3_{\tilde{3}}+\delta A^3_{\tilde{3}}\delta A^2_{\tilde{2}}-\delta A^2_{\tilde{3}}\delta A^3_{\tilde{2}}-\delta A^3_{\tilde{2}}\delta A^2_{\tilde{3}})+\\ \nonumber
    m'^{\; 2}_{\; \tilde{2}}(\delta A^3_{\tilde{3}}\delta A^1_{\tilde{1}}+\delta A^1_{\tilde{1}}\delta A^3_{\tilde{3}}-\delta A^1_{\tilde{3}}\delta A^3_{\tilde{1}}-\delta A^3_{\tilde{1}}\delta A^1_{\tilde{3}})+\\ \nonumber
    m'^{\; 3}_{\; \tilde{3}}(\delta A^1_{\tilde{1}}\delta A^2_{\tilde{2}}+\delta A^2_{\tilde{2}}\delta A^1_{\tilde{1}}-\delta A^2_{\tilde{1}}\delta A^1_{\tilde{2}}-\delta A^1_{\tilde{2}}\delta A^2_{\tilde{1}})
\end{align}

The complication that emerges here is that we have terms in this sum which have coefficients which are dimensionally mass-squared, but the factors of the fields are mixed. That is, rather than terms of the form \(M_{\pi}^2(\delta A^1_{\tilde{1}})^2\) as is usual for a scalar mass-term, we instead have terms like \(M_{\pi}^2(\delta A^1_{\tilde{1}}\delta A^2_{\tilde{2}})\).

It is easy to see that six NG bosons will admit eigenvalues proportional to: $m' \pm {m}'^{1} _{\tilde 1} ,  m' \pm {m}'^{2} _{\tilde 2},m' \pm {m}'^{3} _{\tilde 3}$, namely, $M^2 _{ij} = a (m_i + m_j)$, with $i \ne j$. The other three eigenvalue are obtained by disgonalising the matrix
\begin{equation}
    \begin{pmatrix}
    m' &  {m}'^{1} _{\tilde 1} & {m}'^{2} _{\tilde 2}\\
     {m}'^{1} _{\tilde 1} & m' & {m}'^{3} _{\tilde 3}\\
    {m}'^{2} _{\tilde 2} & {m}'^{3} _{\tilde 3} & m'  
    \end{pmatrix}
\end{equation}

The eigenvalues of the remaining three NG bosons are therefore $M_{\pi}^2 = 2a (m' + \Delta _{1,2,3})$, where $\Delta _{1,2,3} $ are the three roots of the cubic equation
\beq
\Delta ^3 - \Delta ( ({m}'^{1} _{\tilde 1} )^2+ ({m}'^{2} _{\tilde 2})^2 +({m}'^{3} _{\tilde 3})^2  ) - 2  {m}'^{1} _{\tilde 1}\tilde {m}'^{2} _{\tilde 2}\tilde {m}'^{3} _{\tilde 3}=0 \, .
\eeq
In appendix \eqref{cubicappendix} we discuss in detail the solution of this cubic equation and the resulting masses of the pions in various special cases.

The nine massess of the NG bosons obtained by S-duality using the $SO(3)\times SO(3)$ language match the massess obtained by the chiral lagrangian using the $SO(4)$ language, upon the relation $a\sim c$. Thus the radius of the fuzzy sphere is identified with the value of the quark condensate.

\section{Conclusions} 
\label{conclusions}

In this paper we used a non-supersymmetric S-duality to explore the dynamics of a QCD-like theory. Our main result is the derivation of the GMOR relation from the dual magnetic theory. It is interesting to compare the self interactions of the pions that arise from the chiral Lagrangian with those that arise from the magnetic theory. In principle, it is not a hard task: all the terms could be attributed to the self interactions of the scalars in the magnetic theory.

Another open question that deserves further investigation is to identify of the $\eta '$ meson within the magnetic theory. The $\eta '$ transforms together with the nine NG bosons in the coset $U(4)/SO(4)$. In terms of the chiral Lagrangian that we described in section \eqref{chiral} it is the 'missing component' of the two-index symmetric representation of $SO(4)$, namely the $4 \times 4$ unit matrix. 

The presence of massive W bosons in the magnetic theory suggests a 'hidden local symmetry'. It is tempting to identify the W boson with the $\rho$-meson. Similar to the discussion in ref.\cite{Komargodski:2010mc} we expect a rich phenomenolgy, in particular the relation $M_W=gv$ automatically translates into $M^2 _\rho = 2 g^2 _{\rho \pi \pi} f^2_\pi$.  

It will be interesting to further explore the dynamics of other non-supersymmetric gauge theories using S-duality. Some works \cite{Hook:2013vza, armoni:2018ahv,Garcia-Etxebarria:2015hua} have already been carried out in this direction. In particular, based on S-duality in ref.\cite{armoni:2018ahv} it was argued that there is no dynamical symmetry breaking in 3d QCD with matter in the adjoint/symmetric/anti-symmetric representations.

We hope to return to these issues in a future work.

{\bf Acknowledgements} We thank Shigeki Sugimoto for very useful discussions and comments. A.A. wishes to thank YITP, Kyoto, for a warm hospitality, where this work was partly done.

\newpage

\appendix

\section{Appendix: Three-Form Flux Components}

The full, general relation between the quark masses and pion masses under Sugimoto's duality not only allows us to compare the form of the quark mass/pion mass relation with chiral QCD, but will also facilitate comparison between specific cases of quark mass degeneracies and how they affect the distribution of pion masses. In order that we should be able to tune the electric theory fermion masses at will, and to extract the exact relationship between the quark and pion masses under the S-duality, we first must have an explicit expression of the 3-form \(C\).\\
\hfill

To derive an expression for \(C\), we first look at a general term which couples a 3-form, which we will call \(G\), to the fermions in our electric theory.

\begin{equation}
    G_{IJK}(\gamma^{IJK})_{ij}=(m_q)_{ij}
\end{equation}

Where \(m_q\) is a symmetric, \(4\times4\) matrix with real eigenvalues. I,J,K are indices of the \(SO(6)\) space, taking values (\(1,...,6\)). i,j are indices over the \(SO(4)\) space and take values (\(1,...,4\)). We progress with the following procedure.
\begin{equation}
    \gamma^{IJK}=\gamma^{[I}\gamma^J\gamma^{K]}
\end{equation}
\begin{equation} \label{.3}
    G_{IJK}\gamma^{IJK}\gamma^{[I'}\gamma^{J'}\gamma^{K']}=m_q\gamma^{[I'}\gamma^{J'}\gamma^{K']}
\end{equation}\\

Given that \(\gamma^I\) satisfy the Dirac algebra \([\gamma^I,\gamma^J]=2(\gamma^I\gamma^J-\delta^{IJ}\mathbbm{1})\) we can show that,

\begin{equation}
    \Tr(\gamma^{[I}\gamma^J\gamma^{K]}\gamma^{[I'}\gamma^{J'} \gamma^{K']})=
\end{equation}
\begin{equation} \nonumber
    4\delta^{KI'}\delta^{JJ'}\delta^{IK'}-4\delta^{KI'} \delta^{IJ'}\delta^{JK'}+4\delta^{II'}\delta^{KJ'} \delta^{JK'}-4\delta^{II'}\delta^{JJ'}\delta^{KK'}+4 \delta^{JI'}\delta^{IJ'}\delta^{KK'}-4\delta^{JI'} \delta^{KJ'}\delta^{IK'}
\end{equation}\\

Which we substitute into \eqref{.3} and evaluate.
\begin{equation} \label{.5}
    \Tr( G_{IJK}\gamma^{IJK}\gamma^{[I'}\gamma^{J'} \gamma^{K']})=\Tr(m_q\gamma^{[I'}\gamma^{J'} \gamma^{K']})
\end{equation}
\begin{equation} \nonumber
    =G_{IJK}(4\delta^{KI'}\delta^{JJ'}\delta^{IK'}-4\delta^{KI'} \delta^{IJ'}\delta^{JK'}+4\delta^{II'}\delta^{KJ'} \delta^{JK'}-4\delta^{II'}\delta^{JJ'}\delta^{KK'}+4 \delta^{JI'}\delta^{IJ'}\delta^{KK'}-4\delta^{JI'} \delta^{KJ'}\delta^{IK'})
\end{equation}

\begin{equation} \nonumber
    =4G^{K'J'I'}-4G^{J'K'I'}+4G^{I'K'J'}-4G^{I'J'K'}+4 G^{J'I'K'}-4G^{K'I'J'}
\end{equation}
\begin{equation} \nonumber
    =-24G^{[I'J'K']}=\Tr(m_q\gamma^{[I'}\gamma^{J'}\gamma^{K']})
\end{equation}

The reader will notice that we have derived an expression for the components of a vector field, with raised indices, whereas we started this procedure with the aim of finding the components of a three-form, which would have lowered indices. To lower the indices we use the flat metric on the 6-dimensional space transverse to the \textoverline{D3} branes.

As \(G_{IJK}\) are the components of a 3-form, we drop the brackets on the lower indices, which indicate an antisymmetrisation that from here on we will assume tacitly. Therefore, we have,

\begin{equation}
    G_{IJK}=\frac{-1}{24}\Tr(m_q\gamma_{[I}\gamma_J\gamma_{K]})
\end{equation}

\subsection{Anti-Self-Duality of \(C\)}
So far we have an expression for a three-form, that we've called \(G\), which contracts with the anti-symmetric product of three Dirac matrices to give the \(4\times4\) symmetric, real matrix \(M\). This criterion being satisfied is sufficient to support the interpretation of the coupling of \(G\) to the fermions as a sensible mass term. However, the three-form which contracts with the \(\gamma\) triple index used in \cite{Uranga:2004} was anti-self-dual, and so far \(G\) is not. We must therefore go further to assimilate this property into a new three-form, \(C\), which is derived from \(G\).

Let us see what this anti-self-dual property requires:
First, recall that on a Riemannian manifold, the square of the Hodge dual upon a 3-form evaluates to \(-1\), i.e.
\begin{equation} \nonumber
    (\Asterisk_6)^2\omega=-\omega
\end{equation}
Where \(\omega\) is a three-form.\\
\hfill

Uranga gives us the expression for the anti-self-dual three-form \(C\) in terms of an arbitrary three-form, which he calls \(G\), the components of which we have derived explicitly such that its coupling to the fermions is a reasonable mass term. From \cite{Uranga:2004} we have:
\begin{equation}\label{2.13}
    C=(\Asterisk_6G-iG)
\end{equation}
\begin{equation} \nonumber
    \therefore \Asterisk_6C=(-i\Asterisk_6G-G)=-iC
\end{equation}

The Hodge dual convention we follow here is as follows
\begin{equation}
    (\Asterisk\omega)_{IJK}=i\epsilon_{IJK} ^{\quad \;I'J'K'}\omega_{\;I'J'K'}
\end{equation}

Therefore we substitute our expression for the components of  \(G\) into \eqref{2.13} to derive the anti-self-dual components of \(C\)

\begin{equation}
    C_{IJK}=\frac{-1}{48}\Tr(m_q(\epsilon^{IJKI'J'K'}\gamma_{[I'}\gamma_{J'}\gamma_{K']}- i\gamma_{[I}\gamma_J\gamma_{K]}))
\end{equation}

\section{The relation between $m_{q}$ and $m', {m}'^{\; \tilde a}_{\; a}$.} 
\label{relation}

In this work, our approach to investigating Sugimoto's duality in \cite{Sugimoto:2012rt} has been to realise it in a massive theory, and to then employ it to demonstrate the existence a GMOR-like relation between quark and pion masses in the dual theories. To achieve this means to show explicitly the relationship between the matrix \(m_q\), which encodes the quark mass spectrum in the electric theory, and \(m'\) and \({m}'^{\; \tilde a}_{\; a}\) which form the mass-squared term of the pions in the magnetic theory. We can do this in a schematic way very easily, as our derivation of the components for the three-form \(C\) shows it is linear in \(m_q\), and therefore the pion mass-squared is too, i.e. \(M^2_\pi \sim m_q\).

However, to calculate the exact mass-squared spectrum of the pions in terms of the quark masses is a more subtle undertaking. Fundamentally this is a representation theory problem, to appreciate why, recall the symmetry breaking which manifests in both sides of the duality independently:

In the electric theory we initially have a global \(SU(4)\) symmetry, which is broken to \(SO(4)\) when we couple the theory to our three-form flux, that we have derived to confer a sensible mass term to the quarks.

Meanwhile, in the magnetic theory we starts with a global \(SO(6)\) symmetry, which is dynamically broken to \(SO(3) \times SO(3)\) by the fuzzy sphere vacuum.

These symmetry-breaking patterns are equivalent, as we have the well known isomorphism (\(SO(6)\simeq SU(4)\)), and (\(SO(3) \times SO(3) \simeq SO(4)\)) (for which we refer to a publication by Pegoraro \cite{Pegoraro:1975}). However, while these groups are isomorphic they are not the same. They form actions on different spaces, and this difference is significant as we wish to transport eigenvalues between these representation spaces, specifically, the quark masses from the \(SO(4)\) space to the \(SO(3)\times SO(3)\) space, where they constitute the pion mass-squared spectrum.

In \cite{Pegoraro:1975}, Pegoraro discusses two such representation spaces. In one he realises \(\mathbbm{R}^9\) as the linear space of real, traceless, symmetric \(4 \times 4\) matrices, with an action of \(SO(4)\) defined as follows
\begin{equation} \label{B.1}
    SO(4):\hat m \rightarrow O \hat m O  
\end{equation}
Where \(O \in SO(4)\), and \( \hat m\in\) traceless, sym. Mat(\(4,\mathbbm{R})\).

In the other representation, \(\mathbbm{R}^9\) is realised as the linear space of real \(3\times3\) matrices, with the following action of \(SO(3)\times SO(3)\)
\begin{equation}\label{B.2}
    SO(3)\times SO(3):\hat m' \rightarrow O'_1 \hat m' O'_2  
\end{equation}
Where \(O'_1 \in SO(3)_1\),\(O'_2 \in SO(3)_2\) and \(\hat m' \in\) Mat(\(3,\mathbbm{R}\)).

The quark mass matrix \(m_q\) is constrained by our derivation of \(C\) to take values in the space of symmetric, real \(4\times 4\) matrices. Also, it transforms under \(SO(4)\), ergo we can identify \(m_q\) with \(\hat m\) in \eqref{B.1} with the caveat \(m_q\) also contains a traceful component (the singlet of \(SO(4)\)), which we will address separately.

To bring the overall picture together the mass terms our electric and magnetic theories can each be identified as belonging to a representation space of \(SO(4)\) and \(SO(3) \times SO(3)\) respectively. These groups are isomorphic. We also know that as these representation spaces are both isomorphic to \(\mathbbm{R}^9\) there must exist a map between them.

In \cite{Pegoraro:1975}, Pegoraro provides the map we require. For a given matrix \(({m}'^{\; \tilde a}_{\; a})\) in the repn. space of \eqref{B.2}, the corresponding \(m_q\) in \eqref{B.1} has components
\begin{center} \label{B.3}
    \(m_q=\)
    \( \begin{pmatrix}
    {m}'^{\; \tilde{1}}_{\; 1}+{m}'^{\; \tilde{2}}_{\; 2}+{m}'^{\; \tilde{3}}_{\; 3} & {m}'^{\; \tilde{3}}_{\; 2}-{m}'^{\; \tilde{2}}_{\; 3} & {m}'^{\; \tilde{1}}_{\; 3}-{m}'^{\; \tilde{3}}_{\; 1} & {m}'^{\; \tilde{2}}_{\; 1}-{m}'^{\; \tilde{1}}_{\; 2}\\
    {m}'^{\; \tilde{3}}_{\; 2}-{m}'^{\; \tilde{2}}_{\; 3} & {m}'^{\; \tilde{1}}_{\; 1}-{m}'^{\; \tilde{2}}_{\; 2}-{m}'^{\; \tilde{3}}_{\; 3} & {m}'^{\; \tilde{2}}_{\; 1}+{m}'^{\; \tilde{1}}_{\; 2} & {m}'^{\; \tilde{1}}_{\; 3}+\tilde{m}'^{\; \tilde{3}}_{\; 1}\\
    {m}'^{\; \tilde{1}}_{\; 3}-{m}'^{\; \tilde{3}}_{\; 1} & {m}'^{\; \tilde{2}}_{\; 1}+{m}'^{\; \tilde{1}}_{\; 2} & -{m}'^{\; \tilde{1}}_{\; 1}+{m}'^{\; \tilde{2}}_{\; 2}-{m}'^{\; \tilde{3}}_{\; 3} & \tilde{m}'^{\; \tilde{3}}_{\; 2}+\tilde{m}'^{\; \tilde{2}}_{\; 3}\\
    {m}'^{\; \tilde{2}}_{\; 1}-{m}'^{\; \tilde{1}}_{\; 2} & {m}'^{\; \tilde{1}}_{\; 3}+{m}'^{\; \tilde{3}}_{\; 1} & {m}'^{\; \tilde{3}}_{\; 2}+{m}'^{\; \tilde{2}}_{\; 3} & -{m}'^{\; \tilde{1}}_{\; 1}-{m}'^{\; \tilde{2}}_{\; 2}+{m}'^{\; \tilde{3}}_{\; 3}
    \end{pmatrix}\)
\end{center}

Since we only wish to consider matrices \(m_q\) of the form \(m_q=\diag(m_1,m_2,m_3,m_4)\), this simplifies inverting the map. By direct evaluation we see the following
\begin{align} \label{B.4}
    {m}'^{\; \tilde{1}}_{\; 1}=\frac{1}{4}(m_{1}+m_{2}-m_{3}-m_{4})\\ \nonumber
    {m}'^{\; \tilde{2}}_{\; 2}=\frac{1}{4}(m_{1}+m_{3}-m_{2}-m_{4})\\ \nonumber
    {m}'^{\; \tilde{3}}_{\; 3}=\frac{1}{4}(m_{1}+m_{4}-m_{2}-m_{3})
\end{align}

With all other entries zero. In addition the $SO(3)\times SO(3)$ singlet admits
\beq
m' = \frac{1}{4}(m_1+m_2+m_3+m_4) \, .
\eeq

\section{Appendix: properties of the cubic equation and pion masses}
\label{cubicappendix}

In section 4 we derived the pion mass term of the magnetic theory under Sugimoto's S-duality. We stated that six of the pions have masses \(M^2_{ij}=a(m_i+m_j)\) where \(i\ne j\), and that the remaining three pions have masses \(2a(m'+\Delta_{1,2,3})\), where \(\Delta_{1,2,3}\) are the roots of the following polynomial.

\begin{equation} \label{eigenpoly}
    \Delta^3-\Delta((m'^1_{\tilde{1}})^2+(m'^2_{\tilde{2}})^2+(m'^3_{\tilde{3}})^2)-2m'^1_{\tilde{1}}m'^2_{\tilde{2}}m'^3_{\tilde{3}}=0
\end{equation}

This is a depressed cubic equation, which has a set of known solution methods. The method we employ here is Vieta's substitution, which proceeds as follows.

For the general depressed cubic
\begin{equation} \label{Vieta}
    t^3+pt+q=0
\end{equation}

We make the substitution \(t=w-\frac{p}{3w}\), which transforms \eqref{Vieta} to the form
\begin{equation} \label{Vquad1}
    (w^3)^2+q(w^3)-\frac{p^3}{27}=0
\end{equation}

We can solve this quadratic by the standard formula. For \(W\), any non-zero root of the quadratic \eqref{Vquad1}, let \(w_1,w_2,w_3\) be the cube-roots. The roots of the initial cubic \eqref{Vieta} are then \(t_{1,2,3}=w_{1,2,3}-\frac{p}{3w_{1,2,3}}\).\\
\hfill

Applying Vieta's substitution to \eqref{eigenpoly} yields, firstly, the following quadratic

\begin{equation}
    \Omega^2-\Omega(2m'^1_{\tilde{1}}m'^2_{\tilde{2}}m'^3_{\tilde{3}})+\frac{((m'^1_{\tilde{1}})^2+(m'^2_{\tilde{2}})^2+(m'^3_{\tilde{3}})^2)^3}{27}=0
\end{equation}

Which has a root
\begin{equation} \label{eigenquadroot}
    \Omega=m'^1_{\tilde{1}}m'^2_{\tilde{2}}m'^3_{\tilde{3}}+\sqrt{(m'^1_{\tilde{1}}m'^2_{\tilde{2}}m'^3_{\tilde{3}})^2-\frac{((m'^1_{\tilde{1}})^2+(m'^2_{\tilde{2}})^2+(m'^3_{\tilde{3}})^2)^3}{27}}
\end{equation}

Our aim for this paper is, of course, the extraction of GMOR-like relations, which requires that we express our pion masses in terms of the electric-theory quark masses. In section 4 we stated \(m'^1_{\tilde{1}}, m'^2_{\tilde{2}}, m'^3_{\tilde{3}}\) in terms of the quark masses, and repeat here for convenience.

\begin{align} \label{masscoeff}
    m'^1_{\tilde{1}}=\frac{m_1+m_2-m_3-m_4}{4}\\ \nonumber
    m'^2_{\tilde{2}}=\frac{m_1+m_4-m_2-m_3}{4}\\ \nonumber
    m'^3_{\tilde{3}}=\frac{m_1+m_3-m_2-m_4}{4}
\end{align}

When we expand the factors of \(m'^1_{\tilde{1}}, m'^2_{\tilde{2}}, m'^3_{\tilde{3}}\) in \eqref{eigenquadroot} in terms of \(m_1, m_2, m_3, m_4\), we derive a pair of very large polynomials, which we shall label \(Q(m_q),P(m_q)\). For convenience we express the quadratic root \(\Omega\) in terms of these polynomials.

\begin{equation} \label{eigenquadroot2}
    \Omega=\frac{Q(m_q)}{64}+\frac{1}{48\sqrt{3}}\sqrt{P(m_q)}
\end{equation}

\(P(m_q)\) is related to the discriminant of the depressed cubic by a real, negative factor. It is known for cubic polynomials that a positive discriminant implies that the equation has three real, distinct roots. It can be shown that the polynomial \(P(m_q)\) is non-positive for any choice of the quark masses, and therefore the discriminant is non-negative. There are specific cases of quark mass degeneracy which yield a discriminant of zero, and the effect of this in the magnetic theory pion masses will be explored in example calculations. Excluding these special cases however, we are assured that we will always have real pion masses.

To be complete, we provide the full expressions of \(Q(m_q)\) and \(P(m_q)\).

\begin{center}
    \(Q(m_q)=(m_1)^3+(m_2)^3+(m_3)^3+(m_4)^3\)\\
    \(-m_1(m_2)^2-m_1(m_3)^2-m_1(m_4)^2\)\\
    \(-m_2(m_1)^2-m_2(m_3)^2-m_2(m_4)^2\)\\
    \(-m_3(m_1)^2-m_3(m_2)^2-m_3(m_4)^2\)\\
    \(-m_4(m_1)^2-m_4(m_2)^2-m_4(m_3)^2\)\\
    \(+2m_1m_2m_3+2m_1m_2m_4+2m_1m_3m_4+2m_2m_3m_4\)
\end{center}

\newpage

\begin{center}
    \(P(m_q)=-(9(m_1)^4(m_2)^2-9(m_1)^4(m_2)(m_3)-9(m_1)^4(m_2)(m_4)+9(m_1)^4(m_3)^2-9(m_1)^4(m_3)(m_4)\)\\
    \(+9(m_1)^4(m_4)^2-14(m_1)^3(m_2)^3+3(m_1)^3(m_2)^2(m_3)+3(m_1)^3(m_2)^2(m_4)+3(m_1)^3(m_2)(m_3)^2\)\\
    \(+24(m_1)^3(m_2)(m_3)(m_4)+3(m_1)^3(m_2)(m_4)^2-14(m_1)^3(m_3)^3+3(m_1)^3(m_3)^2(m_4)+3(m_1)^3(m_3)(m_4)^2\)\\
    \(-14(m_1)^3(m_4)^3+9(m_1)^2(m_4)^4+3(m_1)^2(m_2)^3(m_3)+3(m_1)^2(m_2)^3(m_4)-3(m_1)^2(m_2)^2(m_3)^2\)\\
    \(-12(m_1)^2(m_2)^2(m_3)(m_4)-3(m_1)^2(m_2)^2(m_4)^2+3(m_1)^2(m_2)(m_3)^3-12(m_1)^2(m_2)(m_3)^2(m_4)\)\\
    \(-12(m_1)^2(m_2)(m_3)(m_4)^2+3(m_1)^2(m_2)(m_4)^4+9(m_1)^2(m_3)^4+3(m_1)^2(m_3)^3(m_4)\)\\
    \(-3(m_1)^2(m_3)^2(m_4)^2+3(m_1)^2(m_3)(m_4)^3+9(m_1)^1(m_4)^4-9(m_1)(m_2)^4(m_3)\)\\
    \(-9(m_1)(m_2)^4(m_4)+3(m_1)(m_2)^3(m_3)^2+24(m_1)(m_2)^3(m_3)(m_4)+3(m_1)(m_2)^3(m_4)^2\)\\
    \(+3(m_1)(m_2)^2(m_3)^3-12(m_1)(m_2)^2(m_3)^2(m_4)-12(m_1)(m_2)^2(m_3)(m_4)^2+3(m_1)(m_2)^2(m_4)^3\)\\
    \(-9(m_1)(m_2)(m_3)^4+24(m_1)(m_2)(m_3)^3(m_4)-12(m_1)(m_2)(m_3)^2(m_4)^2+24(m_1)(m_2)(m_3)(m_4)^3\)\\
    \(-9(m_1)(m_2)(m_4)^4-9(m_1)(m_3)^4(m_4)+3(m_1)(m_3)^3(m_4)^2+3(m_1)(m_3)^2(m_4)^3\)\\
    \(-9(m_1)(m_3)(m_4)^4+9(m_2)^4(m_3)^2+9(m_2)^4(m_3)(m_4)+9(m_2)^4(m_4)^2-14(m_2)^3(m_3)^3\)\\
    \(+3(m_2)^3(m_3)^2(m_4)+3(m_2)^3(m_3)(m_4)^2-14(m_2)^3(m_4)^3+9(m_2)^2(m_3)^4+3(m_2)^2(m_3)^3(m_4)\)\\
    \(-3(m_2)^2(m_3)^2(m_4)^2+3(m_2)^2(m_3)(m_4)^3+9(m_2)^2(m_4)^4-9(m_2)(m_3)^4(m_4)+3(m_2)(m_3)^3(m_4)^2\)\\
    \(+3(m_2)(m_3)^2(m_4)^3-9(m_2)(m_3)(m_4)^4+9(m_3)^4(m_4)^2-14(m_3)^3(m_4)^3+9(m_3)^2(m_4)^4)\)
\end{center}

Note that in the most general case (where \(m_1 \ne m_2 \ne m_3 \ne m_4\)), \(Q(m_q)\) and \(P(m_q)\) cannot be factorized such that the cube-roots of \(\Omega\) and the roots of \eqref{eigenpoly} can be expressed generally and explicitly in linear terms of the quark masses. Therefore, in order to calculate the resulting pion masses explicitly it is necessary to fix the degeneracy of the quark masses a priori.

In terms of the polynomials \(Q(m_q), P(m_q)\), the roots of \eqref{eigenpoly} are given as follows.

\begin{center}
    \(\Delta_{1,2,3}=\sqrt[3]{\frac{Q(m_q)}{64}+\frac{1}{48\sqrt{3}}\sqrt{P(m_q)}}\)\\
   \( +\frac{3(m_1)^2+3(m_2)^2+3(m_3)^2+3(m_4)^2-2m_1m_2-2m_1m_3-2m_1m_4-2m_2m_3-2m_2m_4-2m_3m_4}{48\sqrt[3]{\frac{Q(m_q)}{64}+\frac{1}{48\sqrt{3}}\sqrt{P(m_q)}}}\)
\end{center}

\newpage

\subsection{Special Degeneracy Cases}

\subsubsection{$m_1=m_2=m_3=m_4$:}

For the case of full quark mass degeneracy, we see immediately that \(m'^1_{\tilde{1}}\), \(m'^2_{\tilde{2}}\), \(m'^3_{\tilde{3}}\) all vanish. Therefore \(Q(m_q)\) and \(P(m_q)\) (equivalent to the discriminant of \eqref{eigenpoly}) also vanish, giving trivial roots for \eqref{eigenpoly}, which means the three non-trivial pions receive no shift from \(m'\). To state it explicitly
\begin{equation}
    \Delta_{1,2,3}=0 \nonumber
\end{equation}

\begin{table}[!ht]
\begin{center}
\begin{tabular}{|c|c|}
\hline
 $M_\pi^2$ & Degeneracy \\
\hline
 $2am_1$ & \(9\) \\
\hline
\end{tabular}
\label{Pionmasstable1}
\end{center}
\end{table}

\subsubsection{$m_1 \ne m_2$, $m_2=m_3=m_4$:}

\begin{align*}
    m'^1_{\tilde{1}}=\frac{1}{4}(m_1-m_2)\\
    m'^2_{\tilde{2}}=\frac{1}{4}(m_1-m_2)\\
    m'^3_{\tilde{3}}=\frac{1}{4}(m_1-m_2)
\end{align*}

This reduces \eqref{eigenquadroot} to
\begin{equation} \label{ex2}
    \Omega=\frac{1}{64}(m_1-m_2)^3
\end{equation}
We see again that \(P(m_q)\) has vanished, hence the discriminant of \eqref{eigenpoly} is also zero in this case.\\
\hfill

\eqref{ex2} has cube-roots
\begin{equation} \label{roots1}
    \sqrt[3]{\Omega}=\omega_{1,2,3}=\frac{1}{4}(m_1-m_2),\textrm{ }\frac{\sqrt{3}+i}{8}(m_1-m_2),\textrm{ }\frac{\sqrt{3}-i}{8}(m_1-m_2)
\end{equation}

As stated previously, the cube-roots \(\omega_{1,2,3}\) relate to \(\Delta_{1,2,3}\) as follows

\begin{equation}
    \Delta_{1,2,3}=\omega_{1,2,3}+\frac{(m'^1_{\tilde{1}})^2+(m'^2_{\tilde{2}})^2+(m'^3_{\tilde{3}})^2}{3\omega_{1,2,3}}
\end{equation}

Evaluating this with the roots \eqref{roots1}, we find the following values for the mass shifts

\begin{center}
    \(\Delta_1=\frac{1}{4}(m_1-m_2)+\frac{3(\frac{1}{4}(m_1-m_2))^2}{3(\frac{1}{4}(m_1-m_2))}=\frac{1}{2}(m_1-m_2)\)\\
    
    \(\Delta_2=\frac{\sqrt{3}+i}{8}(m_1-m_2)+\frac{3(\frac{1}{4}(m_1-m_2))^2}{3(\frac{1}{4}(m_1-m_2))(\frac{\sqrt{3}+i}{2})}\)\\
    
    \(=\frac{\sqrt{3}+i}{8}(m_1-m_2)+\frac{\sqrt{3}-i}{8}(m_1-m_2)\)\\
    
    \(=\frac{\sqrt{3}}{4}(m_1-m_2)\)\\
    
    \(\Delta_3=\frac{\sqrt{3}-i}{8}(m_1-m_2)+\frac{3(\frac{1}{4}(m_1-m_2))^2}{3(\frac{1}{4}(m_1-m_2))(\frac{\sqrt{3}-i}{2})}\)\\
    
    \(=\frac{\sqrt{3}-i}{8}(m_1-m_2)+\frac{\sqrt{3}+i}{8}(m_1-m_2)\)\\
    
    \(=\frac{\sqrt{3}}{4}(m_1-m_2)\)
\end{center}

Summarily
\begin{equation}
    \Delta_1=\frac{1}{2}(m_1-m_2),\textrm{ }\Delta_2=\Delta_3=\frac{\sqrt{3}}{4}(m_1-m_2)
\end{equation}

\begin{table}[!ht]
\begin{center}
\begin{tabular}{|c|c|}
\hline
 $M_\pi^2$ & Degeneracy \\
\hline
 $a(m_1+m_2)$ & \(3\) \\
\hline
$2am_2$ & \(3\) \\
\hline
\(a(\frac{3m_1+m_2}{2})\) & \(1\) \\
\hline
\(a\frac{(1+\sqrt{3})m_1+(3-\sqrt{3})m_2}{2}\) & \(2\) \\
\hline
\end{tabular}
\label{Pionmasstable2}
\end{center}
\end{table}

Note that while these masses at first look dubious, they are consistent with the result of the previous degeneracy case. If one takes $m_1=m_2$, the above pion masses reduce appropriately to \(2am_1\) with a degeneracy of 9.

\subsubsection{\((m_1=m_2)\ne(m_3=m_4)\):}

This degeneracy yields the following

\begin{align*}
    m'^1_{\tilde{1}}=\frac{1}{2}(m_1-m_4),\textrm{ }m'^2_{\tilde{2}}=m'^3_{\tilde{3}}=0
\end{align*}

Substituting into \eqref{eigenpoly} we get
\begin{equation}
    \Delta^3-\frac{(m_1-m_4)^2}{4}\Delta=0
\end{equation}

Which by inspection has the solution
\begin{equation}
    \Delta_1=0,\textrm{ }\Delta_2=\frac{1}{2}(m_1-m_4),\textrm{ }\Delta_3=\frac{1}{2}(m_4-m_1)
\end{equation}

This yields the pion spectrum

\begin{table}[!ht]
\begin{center}
\begin{tabular}{|c|c|}
\hline
 $M_\pi^2$ & Degeneracy \\
\hline
 $a(m_1+m_4)$ & \(5\) \\
\hline
$2am_4$ & \(2\) \\
\hline
\(2am_1\) & \(2\) \\
\hline
\end{tabular}
\label{Pionmasstable3}
\end{center}
\end{table}

\newpage
\providecommand{\href}[2]{#2}

\bibliography{bib15}{}
\bibliographystyle{utphys}

\end{document}